\documentstyle[bbm,amsfonts,epsfig,12pt,here]{article}

\newcommand {\eqref} [1] {(\ref {#1})}
\newcommand {\slsh} [1] {\not{\hbox{\kern-2pt${#1}$}}}
\def\drawbox#1#2{\hrule height#2pt
         \hbox{\vrule width#2pt height#1pt \kern#1pt
               \vrule width#2pt}
               \hrule height#2pt}

\def\Asym#1#2{\vcenter{\vbox{\drawbox{#1}{#2}
               \kern-#2pt       % line up boxes
               \drawbox{#1}{#2}}}}

\newcommand {\beq} {\begin{equation}}
\newcommand {\eeq} {\end{equation}}
  \newcommand {\ber}{\begin{eqnarray*}}
  \newcommand {\eer} {\end{eqnarray*}}
\newcommand {\bea}{\begin{eqnarray}}
  \newcommand {\eea} {\end{eqnarray}}

\newcommand{\None}{${\cal N}=1\ $}

\newcommand{\Dslash}{\,{\raise.15ex\hbox{/}\mkern-12mu D}}

\def\Acknowledgements{\bigskip  \bigskip {\begin{center} \begin{large}
              \bf ACKNOWLEDGMENTS \end{large}\end{center}}}

\begin{document}
\begin{titlepage}
\begin{flushright}{CERN-PH-TH/2005-218

SWAT/05/448}
\end{flushright}
\vskip 1cm

\centerline{{\Large \bf Quark Condensate in Massless QCD}}
\centerline{{\Large \bf from Planar Equivalence}}
\vskip 1cm
\centerline{\large Adi Armoni ${}^{a}$, Graham Shore ${}^{a}$ and Gabriele Veneziano 
${}^{b,c}$}
\vskip 0.1cm

\vskip 0.5cm
\centerline{${}^a$ Department of Physics, University of Wales Swansea,}
\centerline{Singleton Park, Swansea, SA2 8PP, UK}
\vskip 0.5cm
\centerline{${}^b$ Theory Division, CERN}
\centerline{CH-1211 Geneva 23, Switzerland}
\vskip 0.5cm
\centerline{${}^c$ Coll\`ege de France, 11 place M. Berthelot, 75005 Paris, France}

\vskip 1cm

\begin{abstract}

A previous estimate of the quark condensate in one-flavour massless QCD from the known value of the gluino condensate in  super Yang-Mills (SYM) theory is extended to $N_f >1$
by considering the large-$N$ limit of an $SU(N)$ gauge theory with one flavour in the antisymmetric representation and $n_f = N_f -1$ extra flavours in the fundamental
representation. We argue that, even at $n_f  \ne 0$, suitably chosen correlators in this theory can be mapped  to those of SYM as $N\rightarrow \infty$.  We give arguments why this correspondence should be particularly good for $N_f =3$ and compare our prediction with available (real and Monte Carlo) data.
\end{abstract}

\end{titlepage}

\section{Introduction}
\label{introduction}

\noindent

There is overwhelming evidence that QCD is the correct
theory of strong interactions. The validity of QCD is confirmed
by comparing with data its predictions for short distance (hard) processes, where
perturbative calculations can be reliably performed.
On the other hand, it is very difficult to carry out reliable analytic
calculations  in the non-perturbative large-distance regime. 
In supersymmetric theories the
situation is much better, due to holomorphy. In particular the gluino condensate \cite{gluinocond}
can be   evaluated {\it
exactly} in \None super Yang-Mills (referred to in the following as SYM theory)

A while ago, a large-$N$ limit
\cite{Armoni:2003gp,Armoni:2003fb} that enabled to
calculate the quark condensate in one-flavour QCD \cite{Armoni:2003yv} was suggested.
The results were summarized in the review paper \cite{Armoni:2004uu} (where
new results and new ideas are included as well).
The idea is to
view the fundamental representation of the $SU(3)$ gauge theory 
as a two-index antisymmetric representation. The next step was to generalize
to $SU(N)$ and then to take the large-$N$ limit. At large $N$, the
theory with an antisymmetric fermion was shown to be equivalent to 
SYM in its bosonic sector. Since the exact value of the gluino
condensate in known, it was possible to copy its value to the non-supersymmetric
theory under consideration. Therefore, up to $1/N$ corrections (of order 30\%)
it was possible to estimate the value of the gluino condensate in one-flavour QCD.

Following this development, several other works that concern the orientifold large-$N$
equivalence were published. In \cite{Sannino:2003xe,Hong:2004td,Feo:2004mr} the emphasis was put on phenomenological aspects and in \cite{DiVecchia:2004ev,Armoni:2005qr} on the relation with string theory.

The purpose of this letter is to generalize the previous analysis to
multi-flavour QCD and in particular to discuss massless three-flavour QCD. The idea is the following. Consider an $SU(N)$ gauge theory with one Dirac fermion in the two index
antisymmetric representation and $n_f$ additional Dirac fermions in the
fundamental representation. We name this theory QCD-OR$'$. For $N=3$, QCD-OR$'$ becomes multi-flavour QCD with $N_f=n_f+1$ fundamental flavours. This model, when $n_f=2$, was considered in the past \cite{Corrigan:1979xf}, with a different motivation.
When $N \rightarrow \infty$ (while $n_f$ is kept fixed) we can neglect the extra 
fundamental flavours inside the loops and the theory becomes again supersymmetric in its
bosonic sector.

The large-$N$ prediction for the quark condensate of multi-flavour QCD coincides, therefore, with the gluino condensate of \None SYM. However, while both one-flavour  and multi-flavour QCD   approach \None SYM at large $N$, the $1/N$ corrections are different in the two cases. Although we cannot fully control these corrections, we will do our best to estimate them.
Moreover, we will present arguments
that, in a suitable sector, three-flavour QCD may be even closer to SYM than one-flavour QCD. 

The manuscript is organized as follows: in section two we will discuss the planar equivalence of QCD-OR$'$ and SYM at large-$N$. In section three we will discuss the issue of how to match the gluino condensate with an appropriate condensate of QCD-OR$'$. In section four we will use the claimed correspondence to calculate the quark
condensate in massless three-flavour QCD and we will compare our prediction with available ``data".

\section{Planar equivalence and multi-flavour QCD}

\noindent

Consider an $SU(N)$ gauge theory with one Dirac fermion $\Psi$ in the two-index
antisymmetric representation and $n_f$ Dirac fermions $\chi_i$ in the
fundamental representation. While at large $N$ the fermions
$\chi_i$ are suppressed, their presence will affect the finite-$N$
theory. According to planar equivalence, at large $N$ a theory with
an antisymmetric fermion is equivalent to \None SYM in a certain 
`common sector'.  Assuming indeed that fundamental matter can be neglected
at large $N$, QCD-OR$'$ theory interpolates between $SU(3)$ with
$N_f=n_f+1$ flavours and \None SYM as we increase $N$ from three to
infinity. Below we present arguments in favour of the equivalence between
large-$N$ SYM and QCD-OR$'$.

The partition function of QCD-OR$'$, after integration over the fermions, reads
\bea 
& &
{\cal Z}_{\rm OR}' = \int \, DA_\mu \exp\left( -S_{\rm YM}[A]\right) \,
\det \left ( i\not\! D \right ) \, ,  \nonumber \\
& & \det \left ( i\not\! D \right ) =
\det ( i\not\! \partial + \not\!\!  A ^a \, (T^a _{anti} \oplus \underbrace { T^a
_{fund} \oplus T^a _{fund} \oplus ... \oplus  T^a _{fund}}_{n_f\,\,\,{\rm times}} )) \,
\label{part2} .
\eea
The fermionic determinant can be expressed in terms of Wilson loops \cite{WL}. In \cite{Armoni:2004ub} such an explicit construction was used to prove that `orientifold field theory' is equivalent to \None SYM at large $N$\footnote{A lattice version of that proof was communicated to us by A. Patella, to appear.}. Let us extend our previous proof to the present case. We first define $\Gamma [A] \equiv \log \det \left ( i\not\! D \right )$.
Next, we use the results of \cite{WL} to express $\Gamma [A]$ in terms of Wilson loops
\bea
\label{wlineint}
 \Gamma [A] &=&
-{1\over 2} \int _0 ^\infty {dT \over T} 
\nonumber\\[3mm]
 &\times&
\int {\cal D} x {\cal D}\psi
\, \exp 
\left\{ -\int _{\epsilon} ^T d\tau \, \left ( {1\over 2} \dot x ^\mu \dot x ^\mu + {1\over
2} \psi ^\mu \dot \psi ^\mu \right )\right\} 
\nonumber \\[3mm]
 &\times &  {\rm Tr }\,
{\cal P}\exp \left\{   i\int _0 ^T d\tau
\,  \left (A_\mu \dot x^\mu -\frac{1}{2} \psi ^\mu F_{\mu \nu} \psi ^\nu
\right ) \right\}  \, ,
\eea
where $A\equiv A^a (T^a _{anti} \oplus \underbrace { T^a
_{fund} \oplus T^a _{fund} \oplus ... \oplus  T^a _{fund}}_{n_f\,\,\,{\rm times}})$ and similarly $F _{\mu \nu} \equiv F^a _{\mu \nu}(T^a _{anti} \oplus \underbrace { T^a
_{fund} \oplus T^a _{fund} \oplus ... \oplus  T^a _{fund}}_{n_f\,\,\,{\rm times}})$. The key ingredient is that a Wilson loop with quarks
in the adjoint/antisymmetric representation can be written as
\bea
& &  
{\rm tr}\, \exp i\int A ^a \, T^a _{adj.} = 2({\rm tr}\, U {\rm tr}\, U^\dagger -1)
\, , \nonumber \\ 
& &
{\rm tr}\, \exp i\int A ^a \, T^a _{anti} = {1\over 2}\left ( ({\rm tr}\, U)^2 -
{\rm tr}\, U^2 \right ) +  {1\over 2} \left ( ({\rm tr}\, U^\dagger)^2 -
{\rm tr}\, (U^\dagger)^2 \right )\, ,
\eea
where ${\rm tr}\,U$ denotes a Wilson loop in the fundamental representation.

Moreover, at large-$N$,
\bea
& & 
 ({\rm tr}\, U\, {\rm tr}\, U^\dagger -1) \rightarrow {\rm tr}\, U {\rm tr}\, U^\dagger  \nonumber \\
& & 
( {\rm tr}\, U)^2 -{\rm tr}\, U^2 + ({\rm tr}\, U^\dagger)^2 -
{\rm tr}\, (U^\dagger)^2  \rightarrow
 ({\rm tr}\, U)^2 + ({\rm tr}\, U^\dagger)^2 \, ,
\label{relations}
\eea
since $({\rm tr}\, U)^2\sim O(N^2)$ while ${\rm tr}\, U^2\sim O(N)$.
Using the above relations \eqref{relations}, one could prove \cite{Armoni:2004ub} the coincidence of the connected part of Wilson loops correlators in large-$N$ SYM and `orientifold field theory',
\beq
  \langle W_{adj.} W_{adj.}...W_{adj.} \rangle _{\rm conn.}=  \langle W_{anti} W_{anti}...W_{anti} \rangle _{\rm conn.} \, \label{coin} .
\eeq
The coincidence of the Wilson loops \eqref{coin} led to the conclusion that the fermionic determinants of SYM and `orientifold field theory', and hence also the partition functions of the two theories, coincide.
 
We can argue that in the present case our previous proof holds by using the following identity
\bea
& &  
{\rm tr}\, \exp i\int A ^a \, (T^a _{anti} \oplus \underbrace { T^a
_{fund} \oplus T^a _{fund} \oplus ... \oplus  T^a _{fund}}_{n_f\,\,\,{\rm times}}) = \nonumber \\ 
& &
 {1\over 2}\left ( ({\rm tr}\, U)^2 -
{\rm tr}\, U^2 \right ) +  {1\over 2} \left ( ({\rm tr}\, U^\dagger)^2 -
{\rm tr}\, (U^\dagger)^2 \right ) + n_f {\rm tr}\, U + n_f {\rm tr}\, U^\dagger \, . 
\label{partOR$'$}
\eea

It is obvious that, as expected, the relation
with the large-$N$ SUSY theory is not spoiled, since ${\rm tr}\,U$  is $O(N)$, 
while $({\rm tr}\, U)^2$ is $O(N^2)$. Namely at large $N$ 
(and $n_f$ fixed),
\beq
 {1\over 2}\left ( ({\rm tr}\, U)^2 -
{\rm tr}\, U^2 \right ) + n_f {\rm tr}\, U  \rightarrow
 {1\over 2} ({\rm tr}\, U)^2 \, ,
\eeq
and therefore, as in \eqref{coin},
\beq
 \langle W_{SYM} W_{SYM}...W_{SYM} \rangle _{\rm conn.}= \langle W_{QCD-OR'} W_{QCD-OR'}...
W_{QCD-OR'}\rangle _{\rm conn.} \, \label{coin2} \, .
\eeq

Moreover, at finite $N$ we can hope for an
improved equivalence with respect to the one-flavour case since both ${\rm tr}\, U^2$ and $2n_f {\rm tr}\, U$ are $O(N)$, and contribute to (\ref{partOR$'$}) with opposite signs. It is interesting to notice that, precisely for $n_f =2$, the first non-trivial contribution of the Wilson loop (the one proportional
to the YM action) is exactly the same in the QCD-OR$'$ and in SYM theory.

In order to relate the gluino condensate of \None Super Yang-Mills 
with the quark condensate of QCD-OR$'$, we need to couple a source to an appropriate
quark bilinear and check for a coincidence at large $N$. The proof of planar equivalence holds also in the presence of such a source. In the next section we discuss the 
precise form of the matching between bilinear fermion operators in SYM and QCD-OR$'$.

\section{Matching correlators in QCD-OR$'$ and SYM}

\noindent

Naively, the presence of extra fermions in the fundamental representation should not
spoil the large-$N$ convergence of QCD-OR$'$ towards SYM. However, the situation
is actually more subtle. There is no dynamical breaking of a continuous chiral symmetry 
in SYM, since its classical $U_R(1)$ symmetry is anomalous. By contrast, QCD-OR$'$, at generic $N$ \footnote{ For $N=3$ the symmetry gets enlarged to the standard $SU(n_f+1)_L\otimes SU(n_f+1)_R \otimes U(1)_V$ while at $N=2$ it becomes $SU(2n_f)$.}, has  
a non-anomalous $SU(n_f)_L\otimes SU(n_f)_R \otimes U(1)_V \otimes U(1)_A$ symmetry 
whose expected dynamical breaking down to $SU(n_f)_V \otimes  U(1)_V $  gives rise to massless Nambu-Golstone (NG)  bosons. How can a theory without a mass gap (QCD-OR$'$) be equivalent to a theory with a mass gap (SYM)?
Moreover, how best can we match condensates in a theory (SYM) with $N$ discrete
vacua with one (QCD-OR$'$) with a continuous vacuum manifold 
($S^1 \times S^3$ for $n_f=2$)?

The answer that we propose is similar, in spirit, to the one already given \cite{Armoni:2003gp, Armoni:2004ub} as a limitation to planar equivalence even at $n_f =0$. It was argued that planar equivalence should only hold in a certain bosonic sector of the theory. In the case at hand we will have to further restrict this bosonic sector by excluding, for instance, external NG bosons or sources coupled to them. If the external particles or sources are decoupled from single  NG-bosons then  such bosons can only be pair-produced and this is, generically, a subleading effect. As a result, the two theories should become equivalent within that `good' sector in the  planar limit\footnote{Even then it is possible that for certain quantities (e.g.~the topological susceptibility) the planar limit ($N\rightarrow \infty$) and the chiral limit  will not commute.}. 
The purpose of this section is to find  (at least part of)  this restricted sector and, in particular, to identify a combination of $\bar \Psi \Psi $ and $\bar \chi \chi$ that coincides with the gluino condensate of \None SYM at large-$N$ and should admit a smooth large-$N$ limit, i.e. one that commutes with sending the masses of the new fermions to zero.
 
Clearly all the non-abelian flavour currents of QCD-OR$'$ are {\it not} in the `good'  sector. Out of the classical $U(1)_A$ currents of QCD-OR$'$ ,
\beq
\label{currents}
J^{(\Psi)}_{\mu 5} = \bar \Psi \gamma _\mu \gamma _5 \Psi \, ,~~~~~~~~~~
J^{(\chi)}_{\mu 5} = \sum_{i=1}^{n_f} \bar \chi_i \gamma _\mu \gamma _5 \chi_i \, , 
\eeq
we can construct the {\em  non-anomalous} conserved current:
\beq
\label{conscurr}
J^{(c)}_{\mu 5} = J^{(\chi)}_{\mu 5}  - {n_f\over N-2}  J^{(\Psi)}_{\mu 5} \, .
\eeq
For $N=3$ and $n_f =2$ this coincides with the usual octet `$\eta$-meson' current of ordinary three-flavour QCD. We will keep
calling $\eta$ the NG boson associated with the spontaneous breaking of the chiral symmetry induced by (\ref{conscurr}). Let us also define the two `decay constants'
\beq
\label{decaycnst}
\langle 0 | J^{(\Psi)}_{\mu 5} |\eta \rangle = iq_\mu F_\Psi \, , ~~~~~~~~~~
 \langle 0 | J^{(\chi)}_{\mu 5} |\eta \rangle = iq_\mu F_\chi \, .
\eeq
The true $\eta$ decay constant $F_\eta$ is obviously given by:
\beq
\label{Feta}
\langle 0 | J^{(c)}_{\mu 5} |\eta \rangle = iq_\mu F_\eta \, , ~~~~~~~~~~
 F_\eta =  F_\chi - {n_f\over N-2} F_\Psi
\eeq
Notice that $F_\Psi$ is $O(N)$ while $F_\chi$ and $F_\eta$ are $O(\sqrt{N})$ as usual.

Now, while the conserved current $J_{\mu 5}^{(c)}$ and thus $F_\eta$ are RG-invariant, 
this is not the case for each individual anomalous current in (\ref{currents})  
and for the corresponding decay constants in (\ref{decaycnst}). Indeed the latter are not only scale-dependent but also mix under changes of the renormalization scale. 
Using RG-invariance of two non-gauge invariant conserved currents, it may be shown that renormalization/mixing take the simple `factorized' form:
\beq
\label{mixing}
J_{\mu 5 R}^{(i)} = J_{\mu 5 B}^{(i)} + a_i \sum_j c_j J_{\mu 5 B}^{(j)} \; , ~~~~~~~~~~
F^{(i)}_R = F^{(i)}_B + a_i \sum_j c_j F^{(j)}_B \; ,
\eeq
where here $i = (\Psi, \chi)$, $a_i$ are the anomaly coefficients ($a_\Psi = 2(N-2), 
a_\chi = 2n_F$), and $c_i$ are some renormalization coefficients that one can compute 
order by order in perturbation theory. (Here, the suffices ${}_R$ and ${}_B$ denote renormalised
and bare quantities respectively.)

Let us consider now a linear combination of the currents (\ref{currents}) that decouples from the $\eta$:
\beq
\label{Jdec}
J_{\mu 5} ^{(d)} = \frac {B} {F_\eta} \left( F_\chi J^{(\Psi)}_{\mu 5} - F_\Psi J^{(\chi)}_{\mu 5}\right) \, 
\eeq
where $B$ is a normalisation constant to be fixed in a moment.
The anomalous current $J_{\mu 5} ^{(d)}$ has two interesting properties:
\begin{itemize}
\item One can easily show, using eqs.(\ref{mixing}), that $J_{\mu 5} ^{(d)}$ is multiplicatively renormalized:
\beq
\label{MR}
J_{\mu 5 R}^{(d)} = (1 + a_i  c_i ) J_{\mu 5 B}^{(d)} \; ,
\eeq
and thus its decoupling is scale-independent.
\item  The anomaly-coefficient for $J_{\mu 5}^{(d)}$ is just $2B$, i.e.
\beq
\label{anom}
\partial^\mu J_{\mu 5}^{(d)} = 2 B Q  \; ,
\eeq
\end{itemize}
As a result, if we fix $B=N$ the decoupled current $J_{\mu 5}^{(d)}$ will have exactly the same anomaly coefficient as the current $J_{\mu 5}^{(\lambda)}$ in SYM theory. (This also ensures that the two-point function $\langle J_{\mu 5}^{(d)} J_{\nu 5}^{(d)}\rangle$ is $O(N^2)$ just like $\langle J_{\mu 5}^{(\lambda)} J_{\nu 5}^{(\lambda)}\rangle$.)
It therefore seems compelling to map these two currents into each other in the large-$N$ limit. With this normalisation, we can rewrite $J_{\mu 5}^{(d)}$ in the useful form
\beq
\label{Jdec2}
J_{\mu 5} ^{(d)} = {N\over N-2} \Bigl(J^{(\Psi)}_{\mu 5} - 
{F_\Psi\over F_\eta}  J^{(c)}_{\mu 5}\Bigr)
\eeq
The overall normalisation factor $(1-{2\over N})^{-1}$ will play a key role below.

Identifying pseudoscalar and scalar operators in the QCD-OR$'$ theory to match with
the components $\bar\lambda \gamma_5 \lambda$ and $\bar\lambda \lambda$ of the chiral
superfield $S^{(\lambda)}$ in SYM is less clear-cut. In particular, the anomalous
Ward identities fix the dependence of the condensate $\langle S^{(\lambda)}\rangle$
on the vacuum angle $\theta$ as $e^{i\theta/N}$, where the factor of $N$ is directly
related to the occurrence of the $N$ discrete vacua in SYM.
A similar analysis applied to QCD-OR$'$, however, shows that the condensate corresponding to
$U = \bar\Psi \Psi + i \bar\Psi\gamma_5 \Psi$ has a $\theta$-dependence~
$\exp\bigl[i {\theta\over N-2} \bigl(1 + {n_f\over N-2}{F_\Psi\over F_\eta}\bigr)\bigr]$ 
while that for $V = \bar\chi \chi + i \bar\chi \gamma_5 \chi$ is~ 
$\exp\bigl[i {\theta\over n_f}\bigl(1 - {F_\chi \over F_\eta}\bigr)\bigr]$. 
The presence of the decay constants in the phase
is unusual and reflects the fact that in QCD-OR$'$ a NG boson couples to an anomalous 
$U(1)$ current. However, recalling the definition (\ref{Feta}) of $F_\eta$ in terms of
$F_\Psi$ and $F_\chi$, we see that if we construct the combination 
$T = U^{N-2} {\rm det}V$ we obtain an operator whose condensate has the correct
$\theta$-dependence, $\langle T\rangle \sim e^{i \theta}$ to match
the one of  $S^{(\lambda)N}$ in SYM (while dimensions only match at $n_f =2$).  
Moreover, we can easily check from the Ward identities that $T$, being a singlet 
under the conserved $U(1)$, decouples from the NG boson, 
$\langle 0|T|\eta\rangle = 0$.

At first sight it appears that this operator has all the desired properties
for  its expectation value to match  $\langle S^{(\lambda)N}\rangle$ of SYM.  However, planar equivalence holds only for diagrams with a fixed number of external legs, and this is not the case for correlation functions of $T$, which involve diagrams with 
$O(N)$ external legs. If we had a large-$N$ estimate of the ratio of $\langle T\rangle$ 
to $\langle S^{(\lambda)N}\rangle$, each one in units of their respective RG-invariant scales, we could use factorization to arrive at the ratio of the bilinear condensates. Unfortunately, we see no way, at present, to proceed along these lines: it would imply  performing a reliable instanton calculation in QCD-OR$'$ , similar to the one that gives the exact condensate in SYM. But then we would have already reached our goal without going through the planar correspondence!  We shall proceed instead in the opposite way: we will  discuss planar equivalence (and $1/N$ corrections to it) directly in terms of bilinear condensates, then use factorization to reconstruct $\langle T\rangle$, and see whether the outcome makes sense in the light of the discussion given above. If this is the case, we shall consider our QCD-OR$'$, SYM condensate relation to be reasonable, and use it at 
$N=3$ to obtain the quark condensate in QCD.

A decoupled pseudoscalar bilinear is easily obtained by expanding $T$ about its VEV:
\beq
P ^{\rm (d)} = C~\biggl[\bar \Psi \gamma _5 \Psi + 
{n_f \over (N-2)} {\langle \bar \Psi \Psi\rangle \over \langle\bar \chi \chi \rangle} 
\bar \chi \gamma _5 \chi \biggr].
\eeq
where $C$ is a normalisation constant which we may take as $1 + O(1/N)$.
$P^{(d)}$ is bilinear in the QCD-OR$'$ fields $\Psi$ and $\chi$ and maps naturally onto the
operator $\bar\lambda \gamma_5 \lambda$ in SYM in the large $N$ limit, where the 
additional term proportional to $\bar\chi \gamma_5 \chi$ is subleading. We can also 
verify, directly from the Ward identities, that it decouples from the NG boson,
$\langle 0|P ^{(d)}|\eta \rangle = 0$.  It therefore satisfies most of the properties we
want in order to match to $\bar\lambda \gamma_5 \lambda$. However, we have to recognise
that this still represents a compromise, since at $O(1/N)$ the $\theta$-dependence of the
corresponding VEV no longer matches that of $\langle S\rangle$, reflecting the 
fundamental difference in the vacuum structure of SYM and QCD-OR$'$ at finite $N$. 

Using $J_{\mu 5}^{(d)}$ and $P^{(d)}$ to match the corresponding operators in SYM,
it is natural to match the SYM condensate\footnote{Notice that from this point on, 
we revert to using Weyl notation for the gluino.}
$\langle \lambda \lambda\rangle \equiv \langle \lambda ^a _ \alpha \lambda ^{a,\alpha} \rangle $ 
with the condensate $\langle S^{(d)}\rangle$
arising in the anomalous Ward identity
\beq
\label{AWI} 
\partial _\mu \langle J_{\mu 5} ^{(d)}~ P ^{(d)} \rangle =  2N \langle Q ~P ^{(d)}
 \rangle  + 2 \langle S^{(d)} \rangle 
\eeq
Recalling the form (\ref{Jdec2}) of the decoupled current, we readily find
\beq
\label{Sdec}
S^{(d)}  = C \left( \frac{N}{N-2}\right) \left[ \bar \Psi \Psi + 
\frac{n_f}{N-2} \frac{F_\Psi}{ F_\eta} 
\left( \bar \Psi \Psi  -  {\langle \bar \Psi \Psi \rangle \over \langle \bar \chi 
\chi \rangle } \bar \chi \chi \right)\right]  \, , 
\eeq
where we stress that the second term in the square brackets, which is the 
variation of $P^{(d)}$ corresponding to the conserved current, necessarily has a 
vanishing VEV. We see, therefore, that although the {\it operator} $S^{(d)}$ is a 
linear combination of $\bar \Psi \Psi$ and $\bar \chi \chi$, 
the appropriate {\it condensate} $\langle S^{(d)}\rangle$ in QCD-OR$'$
to be matched to $\langle \lambda \lambda\rangle$ in SYM is simply 
$\langle \bar\Psi \Psi\rangle$ itself, up to an important normalisation.
The overall factor $(1 - {2\over N})^{-1}$ in (\ref{Sdec}) is just the
normalisation factor defining the decoupled current $J^{(d)}_{\mu 5}$. 
It has a clear physical significance related to the vanishing of the condensate 
$\langle \bar\Psi \Psi\rangle$ at $N=2$, where the antisymmetric representation 
fermion becomes a colour singlet.  Unfortunately, we have no such clear argument 
for the normalisation factor $C$ for $P^{(d)}$. 

Consider therefore, as in \cite{Armoni:2003yv},  the ratio of ratios
\beq
\label{RoR}
K(1/N,n_f) \equiv  
\frac {\langle \bar \Psi \Psi \rangle_{RGI}/\Lambda_{OR'}^3}{ \langle\lambda 
\lambda\rangle/\Lambda_{SYM}^3}
~~=~~
\left(1-{2\over N}\right) {1\over C} \frac {\langle S^{(d)}\rangle_{RGI} / 
\Lambda_{OR'}^3}{ \langle\lambda \lambda \rangle /\Lambda_{SYM}^3} 
\eeq
where planar equivalence and the above identification of $\langle S^{(d)}\rangle$ 
with the gluino condensate $\langle \lambda \lambda\rangle$ ensures that
$K = 1 + O(1/N)$. As in ref.\cite{Armoni:2003yv}, it is convenient to
write
\beq
\label{Kdef}
K(1/N,n_f) = \left(1 - {2\over N}\right) K_*(1/N,n_f)
\eeq
where we explicitly display the normalisation factor arising from $J^{(d)}$
but absorb the unknown normalisation $C$ from $P^{(d)}$ into $K_*$.
We will then argue on more general grounds how the $O(1/N)$ corrections 
to $K_*$ may be constrained.

%Notice that here $ \langle \lambda \lambda \rangle$ is the 
%usual RGI gluino condensate while $\langle S^{(d)} \rangle_{RGI}$ is made RGI 
%through the appropriate power of the 't-Hooft coupling \cite{Armoni:2003yv}.
%Also, as discussed in detail in \cite{Armoni:2003yv}, the gluino condensate 
%is given by:

Before we match the condensates of the two theories we wish to explain our conventions. 
Our slightly unconventional  two-loop $\Lambda$ parameters are defined in terms of the renormalization scale $\mu$ and the 't Hooft coupling $\lambda(\mu) \equiv \frac{\alpha(\mu) N}{2 \pi}$ by:
\beq
\Lambda = \mu (\lambda(\mu))^{-{ \beta _1 \over \beta _0 ^2}} 
\exp\biggl(- {N\over \beta_0 \lambda (\mu)}\biggr)
\eeq
where  $\beta_0$ and $\beta_1$ are the  one and two-loop beta function coefficients, respectively
(for discussion of the subtraction scheme see ref.\cite{Armoni:2003yv}).
We also use
the following standard convention for both QCD-OR$'$ and SYM:
\beq
{\cal L} = -{1\over 2g^2} {\rm tr}\, F^2 + i \bar q \Dslash q \, ,
\eeq
($q$ here can be either $\Psi , \chi$, or $\lambda$) and 
 define a renormalization group invariant quark condensate
as follows
\beq
\langle \bar q q \rangle _{RGI} \equiv \langle 
(\lambda)^{\gamma \over \beta_0} \bar q q \rangle \, ,
\eeq
where  $\gamma$ is  the one-loop anomalous dimension coefficient for the fermion bilinear. Note that for SYM, $\gamma = \beta_0 =3N$. In these conventions \cite{gluinocond}:
\beq
\frac{\langle \lambda \lambda \rangle_{RGI}}{ \Lambda_{SYM}^3} = - \frac{N^2}{2 \pi^2} 
\eeq
Putting all this together, we end up with:
\beq
\label{final}
\frac{\langle \bar{\Psi} \Psi\rangle_{RGI}}{\Lambda_{OR'}^3} =
- \frac{N^2}{2 \pi^2} \Bigl(1 - {2\over N}\Bigr) K_*(1/N,n_f)
\eeq

Before proceeding to estimate $K_*$ let us check whether eq.(\ref{final}) gives 
something reasonable for the corresponding ratio of $\langle T\rangle$ and 
$\langle S^{(\lambda)N}\rangle$. Let us define:
\beq
\label{chiratio}
\frac{\langle \bar{\chi} \chi\rangle_{RGI}}{\Lambda_{OR'}^3} =
 - \frac{N }{2 \pi^2} \tilde{K}(1/N,n_f) \, ,
\eeq
with $\tilde{K}(1/3,n_f) =  K_*(1/3,n_f)$ but where $\tilde{K}(0,n_f)$ need not 
necessarily be 1. Using factorization\footnote{Factorization is exact at large $N$ but it also 
expected to be a good approximation at finite $N$ provided the bilinear 
condensate is dynamically favoured over multifermion condensates.} we would 
predict:
\beq
\frac{\langle T\rangle/  \Lambda_{OR'}^{3(N  + n_f -2)} }{\langle  S^{(\lambda)N}
\rangle/ \Lambda_{SYM}^{3N} } = K_*^{N-2} \tilde{K}^{n_f} N^{n_f} 
\frac{(N-2)^{N-2}}{N^{N+2}} \sim  \frac{K^{N-2} \tilde{K}^{n_f} N^{n_f}}{e^{2}N^2}
\frac{(N-2)!}{N!}
\eeq
Is this result compatible with the assumption $K_*, \tilde{K} \sim 1$ at 
large $N$? We believe that it is. In particular, an $(N-2)!$ is known to originate from 
 distributing the $2N$ 
gluino zero modes among the $2N$ fermions appearing in the correlator 
$\langle S^{(\lambda)N}\rangle$ \cite{Amati}. In the QCD-OR$'$ case a similar argument applied 
to $\langle T\rangle$ would naturally produce an $(N-4)!$ corresponding to the 
smaller number of $\Psi$ zero modes contributing to that correlator. Notice that the number of 
fermionic zero modes is independent of the single-instanton approximation. It depends only
on the topological charge of the gauge configurations contributing to $\langle S^{(\lambda)N}\rangle$
and $\langle T\rangle$.

We finally turn to estimating  $K_*$ and $\tilde{K}$. Up to order $1/N^2$, $K$
has the general form 
\beq
K(1/N, n_f) = 1 ~+~ {{\alpha + \beta n_f}\over N} ~+~ {{\gamma + \delta n_f
+ \epsilon n_f^2 }\over N^2}
\eeq
Implementing the condition that $K$ vanishes at $N=2$ {\it for all} $n_f$
imposes three constraints on the parameters $\alpha, \ldots \epsilon$,
including $\epsilon = 0$. At this order, we can therefore approximate $K_*$ by
\beq
K_*(1/N,n_f)= 1 + {\alpha + 2 \over N} + {\beta n_f \over N} \, ,
\eeq

%Up to order $1/N^2$ for $K$ itself, we can approximate $K_*$ by  
%\beq
% K_*(1/N,n_f)= 1 + {c \over N} + {c' n_f \over N} \, ,
%\eeq
%since higher powers of $n_f$ are necessarily $O(1/N^2)$ and thus 
%give $O(1/N^3)$ contributions to $K$.

In order to estimate the size of $\alpha$ and $\beta$ we use the following argument.
The QCD-OR$'$ theory loses asymptotic freedom and becomes IR free when 
\beq
n_f > \frac{9}{2} N +2 \equiv n_{AF}
\eeq  
It is clear that all condensates should vanish for such large values of $n_f$.
This is implemented in (\ref{final}) by the vanishing of $\Lambda_{OR'}$
when $\beta_0 = 0$ (see next section). 
On the other hand, even in the range with  $n_f < n_{AF}$, the theory, while 
still asymptotically free,  develops a non-trivial  IR fixed point \cite{Banks:1981nn}. 
It is 
reasonable to parametrize the lower limit  $n_{CW}$ of such a `conformal window' 
($n_{CW}< n_f < n_{AF}$) as:
\beq
n_{CW} =  a N +b \, ,
\eeq 
and to assume that all condensates go to zero as $n_f \rightarrow n_{CW}$.
We implement this condition directly in $K_*(1/N,n_f)$.

Within these approximations, we easily find the following unique solution 
for $K_*$ and 
$\tilde{K}$:
\beq
\label{Kcw}
K_* =   \tilde{K} = \Bigl(1-\frac{1}{aN}(n_f -b)\Bigr) \, .
\eeq
which shows clearly that the value of the condensate decreases with an
increasing number of flavours.

Together with the factor $(1 - {2\over N})$ in (\ref{Kdef}), this gives an 
expression for $K$ which implements the two constraints we have on the 
condensate -- vanishing at $N=2$ where the antisymmetric representation 
degenerates to a colour singlet, and the existence of a conformal window -- along 
with planar equivalence, which fixes the large-$N$ limit $K(0,n_f) = 1$.  
For fixed $n_f$, we therefore have information on $K(1/N,n_f)$ at three values 
of $N$. Combining (\ref{Kdef}) with (\ref{Kcw}) fits this `data' with a quadratic 
form, which we may expect to be a good approximation.
 
Expectations\footnote{There is no easy criterion for determining the lower
boundary of the conformal window. However, in ref.\cite{Gardi:1998ch},
Gardi and Grunberg discuss one criterion, based on the analyticity properties of 
the coupling as determined by the beta function, which is simple to estimate. 
With our conventions (see section 4), this condition is $-\beta_0^2/2\beta_1 > 1$.
For QCD with $N_f$ fundamental flavours, this gives $N_{fCW} \simeq 3.2 N$. This 
is higher than some lattice estimates \cite{Iwasaki:2003de}, which give $N_{fCW} 
\simeq 7$ for $N=3$. Nevertheless, we may use the analyticity criterion to get at 
least a rough estimate of $n_{CW}$ in QCD-OR$'$. For QCD-OR$'$, with one fermion 
in the antisymmetric representation and $n_f$ fundamentals, the beta 
function coefficients are
\beq
\beta_0 = 3N - {2\over3}(n_f-2)~~~~~~~~~~~~~~
\beta_1 = 3N^2 + 2N - {3\over N} - \left({13\over6}N-{1\over2N}\right)(n_f-2)
\eeq
We can then show that the condition $-2\beta_1 \simeq \beta_0^2$ is 
approximately satisfied for $N \ge 3$ by $n_f = aN + b$ with
$a \simeq 2$, $b \simeq 2.6$.}
for $a$ and $b$ are around $2\pm 1$ \cite{Gardi:1998ch}. Inserting 
$a=b=2$ in (\ref{Kcw}), our final result for the 
three-flavour QCD condensate, $N=3, n_f =2$, becomes:
\beq
\label{N=3n=2}
\frac{\langle \bar{\Psi} \Psi\rangle_{RGI}}{\Lambda_{OR'}^3} =
 - \frac{9}{2 \pi^2} K(1/3, 2) \simeq - 0.15 \pm 0.05
\eeq
The error arises from the uncertainty involved in the truncation of $K$
and the values of $a$, $b$ determining the conformal window. Notice, however, that
the latter is minimised for $n_f \simeq b \simeq 2$. We conservatively estimate 
the total error as 30\%.

In the following section, after giving some more arguments why the case $n_f=2$ 
should be particularly favourable, we turn to a discussion of the comparison of 
our prediction with real and Monte-Carlo data.

\section{Quark condensate in massless three-flavour QCD}

\noindent 

So far we have considered QCD-OR$'$ for general $n_f$, which reduces to 
multi-flavour massless QCD at $N=3$. We now specialise to $n_f = 2$ and 
estimate the quark condensate in three-flavour QCD.
We restrict ourselves to three flavours, since it is both the most interesting
case from the phenomenological point of view and since our approximation is  
expected to be optimal when $n_f=2$. The reason is the following: as we shall see,
the calculation of the condensate is sensitive to the value of the one and 
two-loop beta function coefficients. For  $n_f=2$ the one-loop beta function of 
QCD-OR$'$ is $\beta _0=3N$, exactly as in \None SYM. The two-loop beta function
is $\beta _1=3N^2(1 + 2/(3N) - 1/N^3)$, which is very close to the value of
the SUSY theory ($\beta _1 = 3N^2$). The underlying reason is as follows: the 
Dynkin index of a certain representation reflects the number of degrees of freedom
that run in the vacuum polarization loop. For the adjoint it is $N$,
while for the antisymmetric representation it is ${1\over 2}(N-2)$ and
for the fundamental representation it is ${1\over 2}$. Since we have
a Dirac fermion in the antisymmetric and two Dirac fermions in the
fundamental we get $(N-2) +2=N$. In this sense, the physical degrees of freedom 
in our model are the same as those of an adjoint Majorana fermion. Similarly, we 
saw in section 2 that the expectation value of a Wilson loop in QCD-OR$'$ might 
approach the expectation value of a Wilson loop in SYM with rather mild $1/N$ 
corrections when $n_f=2$.

Let us proceed to the calculation of the quark condensate. Our conclusion from 
the last section is that for $SU(3)$,
\beq
\langle \bar{\Psi} \Psi \rangle _{RGI} \, \simeq -\frac{3}{2\pi^2} 
\Lambda_{OR'}^3 \label{orcond}
\eeq     
In order to compare the value of the condensate to lattice predictions, it is 
useful to write \eqref{orcond} at an arbitrary scale $\mu$.  Using our previously introduced 
definitions gives:
\beq
\langle \bar \Psi \Psi \rangle_\mu = - {3 \over 2\pi^2} \mu ^ 3 
(\lambda(\mu))^{-{\gamma \over \beta _0} - {3\beta _1 \over \beta _0 ^2}} 
\exp\biggl(- {9\over \beta_0 \lambda (\mu)}\biggr) .
\label{condensate}
\eeq
 For three-flavour QCD the relevant values are
$\beta_0= 9,\, \beta_1=32,\, \gamma =4$. 

\begin{figure}[H]
\begin{center}
\mbox{\kern-0.5cm
\epsfig{file=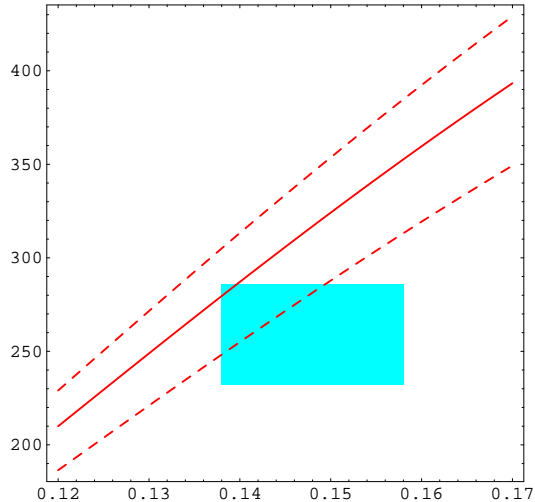,width=7.0true cm,angle=0}}
\end{center}
\caption{The quark condensate expressed as $-(y~{\rm MeV})^3$ as a function of 
the 't Hooft coupling $\lambda$. The $\pm 1\sigma$ range of the coupling, 
$0.138 < \lambda < 0.158$ and the lattice estimate 
$-(259 \pm 27~{\rm MeV})^3$ define the shaded region.}
\label{fig}
\end{figure}

In order to calculate the condensate \eqref{condensate} we need to
know the value of the 't Hooft coupling $\lambda$ at a scale $\mu$
that we choose to be $\mu = 2\,{\rm GeV}$. 
The Particle Data Group \cite{PDG} quotes 
$\alpha_s(2 {\rm~GeV}) = 0.31 \pm 0.01$
which corresponds to $\lambda (2~{\rm GeV}) = 0.148 \pm 0.010$.
We therefore choose to plot the function
\beq
 \frac{\langle \bar \Psi \Psi \rangle_ {\rm 2\, GeV}}{{\rm GeV}^3} = 
- {3 \over 2\pi^2} 2 ^ 3
\lambda ^{- {44
\over 27} } \exp \biggl(- { 1\over \lambda }\biggr) ,
\label{value}
\eeq
in a range of $\lambda$. The results are presented in fig.\eqref{fig}.

The solid line in fig.\ref{fig} shows the condensate  
$\langle \bar \Psi \Psi \rangle_ {\rm 2\, GeV}$ evaluated from (\ref{value})
as a function of the 't Hooft coupling $\lambda(2 {\rm~GeV})$.
With the central value $\lambda = 0.148$, we find
\beq
\langle \bar \Psi \Psi \rangle_ {\rm 2\, GeV} ~~=~~ -(317 ~\pm~ 30~\pm~36~
{\rm MeV})^3
\label{cond}
\eeq 
Here, we have included an error $\pm 30\%$ in the condensate due to the 
uncertainties in estimating $K$ discussed in the previous section. This is
indicated in the plot by the dashed lines.  The second error shown in
(\ref{cond}) reflects the $\pm 1\sigma$ range of the experimentally determined
coupling. 

We can compare this result with estimates from lattice QCD. Up to now,
these have mostly been performed in the quenched approximation with overlap
fermions, with the condensate either evaluated directly or deduced from
mass ratios using the Dashen--Gell-Mann--Oakes--Renner (DGMOR) relation.
Two recent results in this category are those of Giusti {\it et al.}
\cite{Giusti:2001pk} who find $-(267 \pm 16~{\rm MeV})^3$
and Wennekers and Wittig \cite{Wennekers:2005wa} who quote
$-(285 \pm 9~{\rm MeV})^3$. A recent analysis using dynamical fermions
has been given by McNeile \cite{McNeile:2005pd} using MILC data for mass 
ratios together with the DGMOR relation. He finds $-(259 \pm 27~{\rm MeV})^3$. 
(Ref.\cite{McNeile:2005pd} also contains a useful compendium of previous 
lattice estimates.) This is the result shown in the shaded band in 
fig.\ref{fig}, extended across the $1 \sigma$ range of $\lambda(2~{\rm GeV})$.

It is apparent from these comparisons that our method tends to slightly
over-estimate the condensate compared with the lattice evaluations, at least 
within the $1 \sigma$ range of $\alpha_s(2~{\rm GeV})$. A lower coupling would 
obviously ease the agreement. It is also clear that both methods still have 
significant uncertainties. However, bearing this in mind, the degree of 
agreement is extremely encouraging, especially when one remembers that we 
are calculating a non-perturbative quantity in QCD from first principles 
using entirely analytic methods. In conclusion, we find this result very 
promising for the programme of exploiting our knowledge of exact results in 
supersymmetric gauge theories to calculate non-perturbative quantities
in QCD itself.

\vfill\eject
  
\Acknowledgements

A.A. and G.V. would like to thank L. Del-Debbio for very useful discussions.
We thank M. Shifman  for collaboration in
the early stages of this work and for many fruitful discussions and comments. A.A. is supported by the PPARC advanced 
fellowship award.

\end{document}